\documentclass{epl}

\title{
Bistable anchoring of nematics on rough substrates
}
\shorttitle{Bistable anchoring}

\author{{F. Schmid\inst{1} \and D. Cheung\inst{2}}}
\institute{
  \inst{1} Fakult\"at f\"ur Physik, Universit\"at Bielefeld, 
D-33615 Bielefeld, Germany \\
  \inst{2} Dept. of Physics, University of Warwick, Coventry, CV4 7AL, UK 
}


\newcommand{\dir}{Figs}

\newcommand{\QQ}{\mbox{$\mathbf{Q}$}}
\newcommand{\rr}{\mbox{$\mathbf{r}$}}
\newcommand{\nn}{\mbox{$\mathbf{n}$}}

\newcommand{\Tr}{\mbox{Tr}}

\newcommand{\ie}{\mbox{{\em i.e., }}}
\newcommand{\eg}{\mbox{{\em e.g., }}}

\begin{document}

\maketitle 

\begin{abstract}
We analyze the interplay between wetting and anchoring of nematic liquid crystals 
on disordering, \eg \ rough substrates in the framework of the Landau-de Gennes 
theory, in situations of competing homeotropic and planar easy axes on the 
substrate and the nematic-isotropic (NI) interface. The phase diagram for 
azimuthally symmetric substrates is calculated. We identify two regimes - a
strongly coupled regime, where the wetting transition coincides with an anchoring
transition, and a weakly coupled regime, where the two are separated. The
anchoring transition is first order and switches between homeotropic 
and planar anchoring. The two competing orientations are metastable over a 
broad parameter range. Hence such surfaces can be used to generate bistable surfaces.
\end{abstract}

%xxxxxxxxxxxxxxxxxxxxxxxxxxxxxxxxxxxxxxxxxxxxxxxxxxxxxxxxxxxxxxxxxxxxxxxxxxxxxx
% INTRODUCTION

Nematic liquid crystals are fluids of aligned particles with a preferred 
orientation~\cite{degennes}. In the bulk, the direction of alignment is 
arbitrary, but surfaces break the isotropy of space and tend to align 
nematic fluids. This phenomenon, called surface anchoring, is of both 
fundamental and industrial interest, \eg in liquid display 
technology~\cite{cognard,jerome}. From the practical point of view, it 
is useful to have surfaces which favor two orientations, such that 
one can effortlessly switch from one to another. Various strategies have 
been proposed to obtain such bistable 
surfaces~\cite{jerome2,monkade,nobili,alexe,lange,brown,harnau}. 
Here, we analyze a mechanism that generates bistability on rough 
substrates.

Experiments~\cite{barberi,papanek,wen}, theory~\cite{barbero}, as well 
as simulations~\cite{cheung1,cheung2} have shown that rough or 
nanostructured substrates can reduce the order at the surface and even 
depress the NI transition in confined systems. If the 
disordering effect of a substrate is strong enough to nucleate a 
surface layer of strongly reduced order, the adjacent nematic 
fluid is oriented by that layer rather than by the bare substrate~\cite{papanek}, 
and the direction of alignment may change. 
For example, Simoni \etal \ have observed experimentally that 
a porous polymeric substrate imprints a different orientation on a 
nematic film than a flat substrate made of the same material~\cite{simoni}. 
In their system, the alignment was homeotropic on the flat substrate, 
but planar on the porous substrate. This effect can be understood quite 
naturally if one assumes that the substrate favors homeotropic alignment, 
and the NI interface induces planar alignment. 

The question is whether the effect can be used to generate bistable surfaces. 
Rodriguez-Ponce \etal\cite{rodriguez} have carried out density functional
calculations for a model liquid crystal on a disordering substrate in a similar 
situation of competing, planar and homeotropic, anchoring axes. Encouragingly, 
they find a first order anchoring transition between planar and homeotropic 
anchoring. The anchoring transition seems related to wetting, but the exact nature 
of the relation is not clear: The system also exhibits wetting transitions 
that are not connected to an anchoring transition, and even reentrant wetting.
In the present work, we present an analysis of the problem in the general
framework of the Landau-de Gennes theory. We clarify in detail 
the relation between the anchoring transition and the wetting transition,
and identify different regimes of weak and strong coupling, which establish
the context for the findings of Rodriguez \etal. We find that the anchoring 
transition is first order, and that there exists a broad intermediate 
regime, where both anchoring orientations are at least metastable 
and the system can easily switch from one to the other. Hence these surfaces 
should be suitable candidates for bistable surfaces. 

The fact that competing easy axes may influence wetting phenomena has first 
been pointed out for {\em nematic} wetting layers by Sullivan and 
Lipowsky~\cite{sullivan}. Director distortions in nematic wetting 
layer may lead to long-range interactions between the substrate and the 
NI interface. The consequences for the wetting phase behavior have been analyzed 
within the Landau-de Gennes theory by Braun \etal~\cite{braun}, using a 
formalism originally developed by Sen and Sullivan~\cite{sen}. Within the same 
framework, Teixeira \etal~\cite{teixeira} have studied a system with an
isotropic wetting layer. This work is related to ours, but the system is much 
more complicated. Already the bare substrate free energy has two competing 
contributions: A term which is linear in the order parameter and favors surface 
order and homeotropic anchoring, and a quadratic disordering term which 
favors {\em conical} anchoring. As a result, a continuous surface driven 
anchoring transition from conical to homeotropic was found for a set 
of parameters with strong linear term, and a first order wetting driven 
anchoring transition from conical to planar for a second set with weak linear 
term. Here, we shall focus on the simpler case of a purely disordering 
substrate that favors homeotropic anchoring.

%
%xxxxxxxxxxxxxxxxxxxxxxxxxxxxxxxxxxxxxxxxxxxxxxxxxxxxxxxxxxxxxxxxxxxxxxxxxxxxxx
% THEORY

Our starting point is the Landau-de Gennes theory of nematic liquid crystals,
which is based on a free energy expansion in powers of a symmetric and traceless
$(3 \times 3)$ order tensor field $\QQ(\rr)$. The leading terms contributing
to the bulk free energy are~\cite{degennes} 
\begin{equation}
\nonumber
F_{\rm bulk} = \int \!\!\! d^3r \Big\{ \frac{A}{2} \Tr(\QQ^2) 
\! + \! \frac{B}{3} \Tr(\QQ^3) 
\! + \! \frac{C}{4} \Tr(\QQ^2)^2 
%\\
%&&  \qquad
+\: \frac{L_1}{2} \partial_i Q_{jk} \partial_i Q_{jk}
+ \frac{L_2}{2} \partial_i Q_{ij} \partial_k Q_{kj} \Big\}.
\label{eq:landau_q}
\end{equation}
For simplicity, we neglect the biaxiality and approximate the order tensor 
by~\cite{priestley}
$Q_{ij}(\rr) = \frac{1}{2} S(\rr) ( 3 n_i(\rr) n_j(\rr) - \delta_{ij})$,
where $S(\rr)$ is the local nematic order parameter, and $\nn(\rr)$
the director, a vector of length unity describing the local direction 
of alignment. This is justified by the fact that homeotropically
orienting surfaces do not induce biaxiality, and that the biaxiality
induced by the NI interface is small~\cite{popa-nita,chen}. Furthermore, 
we introduce the ``natural'' units $\hat{S} = - {2 B}/{9 C}$,
$\hat{\xi} = 2\sqrt{{(L_1 + L_2/6)}/{3C}} \: {\hat{S}^{-1}}$,
and $\hat{\epsilon} = ({3C}/{16}) \cdot \hat{S}^4 \hat{\xi}^3$.
The quantity $\hat{S}$ is the value of the order parameter in the nematic phase at
coexistence, $\hat{\xi}$ is the minimum width of a planar NI interface at 
coexistence, and $\hat{\epsilon}/\hat{\xi}^2$ the corresponding 
interfacial tension. In the following, all order parameters, lengths,
and energies, shall be rescaled by these units. This leaves us with two 
dimensionless parameters,
\begin{equation}
t = \frac{1}{4} A  \frac{\hat{S}^2 \hat{\xi}^3}{\hat{\epsilon}} - 1,
\qquad \mbox{and} \qquad
\alpha = \frac{1}{2} \: \frac{L_2}{(L_1 + L_2/6)}.
\end{equation}
The parameter $t$ is proportional to the distance to NI coexistence in the 
phase diagram - \ie \ the temperature distance ($t \propto (T-T_{NI})$)
in thermotropic liquid crystals, or the chemical potential distance
($t \propto (\mu-\mu_{NI})/T_{NI}$) in lyotropic liquid crystals.
The parameter $\alpha$ characterizes the anchoring strength of a planar 
NI interface. At $\alpha > 0$, the interface favors parallel, 
planar alignment, and at $\alpha < 0$, it favors perpendicular, homeotropic 
alignment.

The resulting rescaled bulk free energy takes the form
\begin{eqnarray}
\label{eq:landau}
\lefteqn{ 
F_{\rm bulk} = 3 \int d^3r \: \{ f + g_1 + g_2 \} \quad \mbox{with} \quad
f = S^2 ((S - 1)^2 + t), } \qquad \qquad 
\\ \nonumber
g_1 &=& \Big( (\nabla \cdot S)^2 + \alpha (\nn \cdot \nabla S)^2 \Big) +
          4 \alpha S \Big( (\nabla \cdot \nn) (\nn \cdot \nabla S)
          + \frac{1}{2} (\nn \times \nabla \times \nn) (\nabla S) \Big),
\\ \nonumber
g_2 &=& S^2 \Big(
(3 + 2 \alpha) (\nabla \nn)^2
+ (3 - \alpha) (\nn \cdot \nabla \times \nn)^2 
+ \: (3 + 2 \alpha)(\nn \times \nabla \times \nn)^2
\Big).
\end{eqnarray}
The first term, $f(S)$, describes the free energy density of a homogeneous
system, the middle term accounts for the effect of order parameter variations,
and the last term corresponds to the Frank elastic energy of a nematic
phase with spatially varying director~\cite{degennes}. This term, $g_2$, allows
to relate the parameter $\alpha$ to the experimentally accessible Frank elastic
constants $K_i$ of a material. For example, for MBBA, one has
$K_3/K_2=(3+2\alpha)/(3-\alpha) \approx 3$, \ie $\alpha \approx 1.2$. 

Next, we must determine the appropriate surface free energy.
An isotropic surface introduces only one symmetry breaking vector, the 
surface normal $\nn_0$. This vector can be combined with $\QQ$ to construct 
the surface energy terms that are compatible with the symmetry of the 
system~\cite{sluckin}. The linear order term in $\QQ$, 
$ \nn_0 \QQ \nn_0 = \frac{3}{2} S (\nn \nn_0)^2 - \frac{1}{2} S$,
favors nematic order, \ie a surface free energy containing such a term
will always be minimized by a nonzero value of $S$. 
Hence this term must vanish close to a truly disordering surface. 
To quadratic order in $\QQ$, one obtains three terms, from which one can
construct the general expression
\begin{equation}
\label{eq:surf}
F_{\rm surf} = \int d^2r \: f_{\rm surf}
\quad \mbox{with} \quad
f_{\rm surf} = W S^2(1 + \beta n_{\parallel}^2 + \gamma n_{\parallel}^4)
\quad \mbox{and} \quad n_{\parallel}^2 := 1-(\nn \nn_0)^2.
\end{equation} 
At $W>0$ and $\beta>0$, the surface favors $S=0$ (disorder) and $n_{\parallel}=0$ 
(homeotropic alignment). The parameter $W > 0$ measures the disordering effect of 
the substrate, \ie it's roughness, and the parameter $\beta >0$ characterizes it's 
orienting strength. For $\gamma < -\beta/2$, the surface has an additional
preference for planar anchoring. In the following, we 
shall assume $\gamma = 0$ for simplicity. It is worth noting that the form 
(\ref{eq:surf}) of $f_{\rm surf}$ already implies that the wetting transition 
{\em must} be second order, for symmetry reasons, unless it is coupled with 
an anchoring transition~\cite{fn1}. 

The total free energy is given by $F = F_{\rm bulk} + F_{\rm surf}$, 
with $F_{\rm bulk}$ and $F_{\rm surf}$ given by Eqs.~(\ref{eq:landau}) 
and (\ref{eq:surf}). Our task is to minimize this functional with respect 
to the profiles $S(\rr)$ and $\nn(\rr)$. We take the surface to lie 
in the $(xy)$ plane, hence we can assume that the profiles vary only in the 
$z$ direction. After parametrizing the director $\nn$ as 
$\nn = (n_{\parallel} \cos \phi, n_{\parallel} \sin \phi, n_z)$ with 
$n_{\parallel}^2 + n_z^2 = 1$, one checks easily that $F$ as a function of 
$n_z$ and $\phi$ is minimized by $d \phi/dz \equiv 0$: The director does not 
vary in the azimuthal direction.  Hence we are left with two profiles, 
$S(z)$ and $n_z(z)$, which have to be determined such that they
minimize the total free energy.

We will now sketch a method that allows to solve this and similar 
problems very efficiently. We divide the order parameter $S(z)$ 
profile in piecewise monotonic parts. (In our case, the whole 
profile was monotonically increasing). For each part, we rewrite the director 
profile as a function of the order parameter $S$, $n_z(z) \equiv n_z(S)$.
After introducing new variables $q = \ln (S)$ and $\psi = 2 \: \mbox{arcsin} (n_z)$
for convenience, the bulk free energy (\ref{eq:landau}) per surface area $A$ 
can be written
\begin{eqnarray}
\frac{F_{\rm bulk}}{A} &=& 3 \int_0^{\infty} \!\!\! dz\: 
\Big\{ f(e^q) + (\frac{dq}{dz})^2 \: e^{2 q} \: 
\Phi^2(\psi,\frac{d\psi}{dq}) \Big\} 
\\
\label{eq:phi}
&& 
\mbox{with} \qquad
\Phi^2(\psi,\psi') = 1 + \frac{\alpha}{2} (1-\cos \psi)
+ \frac{\alpha}{2} \psi' \sin \psi 
+ \frac{3 + 2 \alpha}{4} \psi'^2.
\nonumber
\end{eqnarray}
It is first minimized with respect to $q(z)$ for given $\psi(q)$.  
The Euler-Lagrange equations yield the integration constant
$f(e^q)-(dq/dz)^2 e^{2 q} \Phi^2 = \mbox{const.} = f_{\infty}$,
where $f_{\infty} = f(S_{\rm nematic})$ is the free energy density 
in the homogeneous nematic bulk. This can be used to derive
an expression for $dq/dz$, which can be inserted into the total 
free energy, yielding
\begin{equation}
\label{eq:ftotal}
F/A = V f_{\infty}/A + 6 \int_{q_0}^{q_{\infty}} \!\!\! dq \: e^q \: 
\sqrt{f(e^q) - f_{\infty}} \: \Phi(\psi,\psi')
+ f_{\rm surf}
\end{equation}
with $f_{\rm surf} =  W e^{2 q_0} (1+ \beta \cos^2(\psi_0/2)
+ \gamma \cos^4(\psi_0/2))$. 
The index $\infty$ stands for the bulk, the index $0$ for 
the surface, and $V$ is the total volume of the system.
The free energy (\ref{eq:ftotal}) can now be minimized 
with respect to $\psi(q)$. A variational treatment 
yields the Euler-Lagrange equation 
\begin{equation}
\label{eq:euler}
\psi'' \frac{\partial^2 \Phi}{\partial \psi'^2}
= \frac{\partial \Phi}{\partial \psi}
- \frac{\partial^2 \Phi}{\partial \psi \partial \psi'} \psi'
- \Big(1 + \frac{1}{2} \frac{d}{dq} \ln(f(e^q)-f_{\infty}) \Big)
\frac{\partial \Phi}{\partial \psi'},
\end{equation}
\begin{eqnarray}
\mbox{with boundary conditions} 
\quad \qquad 
\label{eq:boundaryb}
\frac{\partial \Phi}{\partial \psi'}\Big|_{q_{\infty}} = 0,
&\; &
q_{\infty} = \ln (S_{\rm nematic}),
\\
\mbox{and} \quad
\label{eq:boundary0}
6 e^{q_0} \sqrt{f(e^q_0)-f_{\infty}} \:
\frac{\partial \Phi}{\partial \psi'}\Big|_{q_0} = 
\frac{\partial f_{\rm surf}}{\partial \psi}\Big|_{q_0},
& \; &
6 e^{q_0} \sqrt{f(e^{q_0}) - f_{\infty}}\: \Phi(\psi_0,\psi_0')
= \frac{\partial f_{\rm surf}}{\partial q}\Big|_{\psi_0}.
\end{eqnarray}
Eq.~(\ref{eq:boundaryb}) (left) was obtained by minimizing (\ref{eq:ftotal}) 
for arbitrary upper integration limit $q_{\rm max} < q_{\infty}$, and then 
taking the limit $q_{\rm max} \to q_{\infty}$. One easily checks that both 
$\psi \equiv 0$ and $\psi \equiv \pi$ are solutions of 
Eqns.~(\ref{eq:euler})-(\ref{eq:boundary0}). To calculate the anchoring
potential for arbitrary anchoring angle $\theta$, we fix 
$\psi_{\infty}=\pi - 2 \theta$, calculate $\psi_{\infty}'$ from 
Eq.~(\ref{eq:boundaryb}), and perform a straightforward integration of 
Eq.~(\ref{eq:euler}), starting at $q_{\infty}$ and stopping as soon as 
Eq.~(\ref{eq:boundary0}) (right) is fulfilled. The resulting total free energy 
per area $F/A \equiv \Sigma$ gives the anchoring potential. At the
extrema of $\Sigma$, the remaining boundary condition (\ref{eq:boundary0}) 
(left) is automatically fulfilled.

\begin{figure}
\oneimage[scale=0.28]{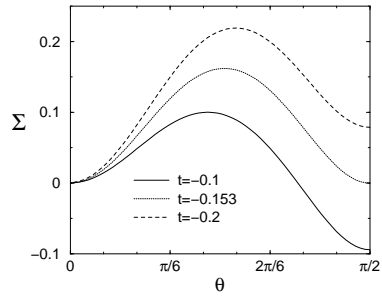}
\caption{Examples of effective anchoring potentials $\Sigma$ as a function 
of the anchoring angle $\theta$ between the director and the surface normal.
The parameters are $W=2.5, \alpha = 1.2, \beta = 1.$, and $t$ as indicated.
}
\label{fig:potentials}
\end{figure}

\begin{figure}
\twoimages[scale=0.28]{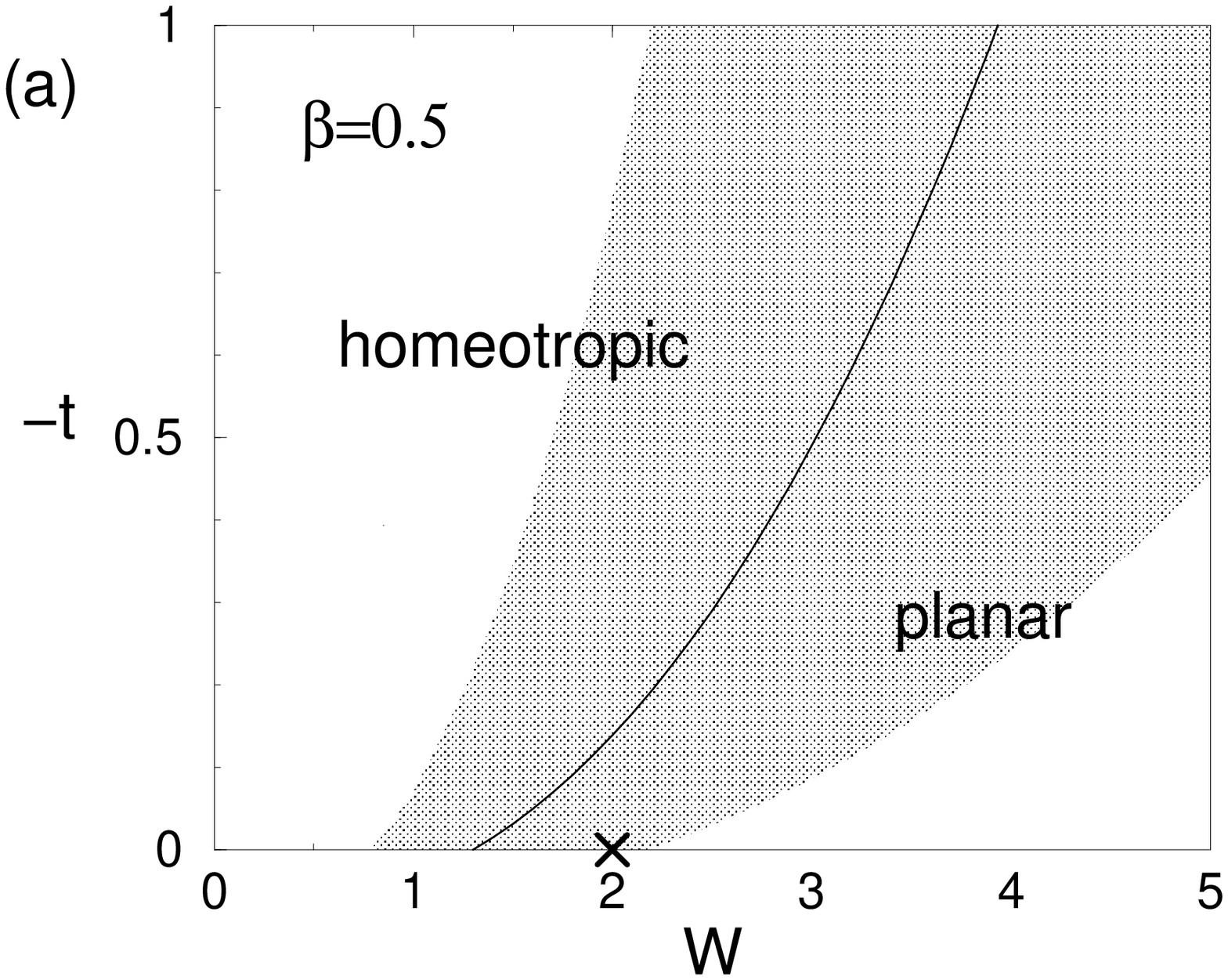}{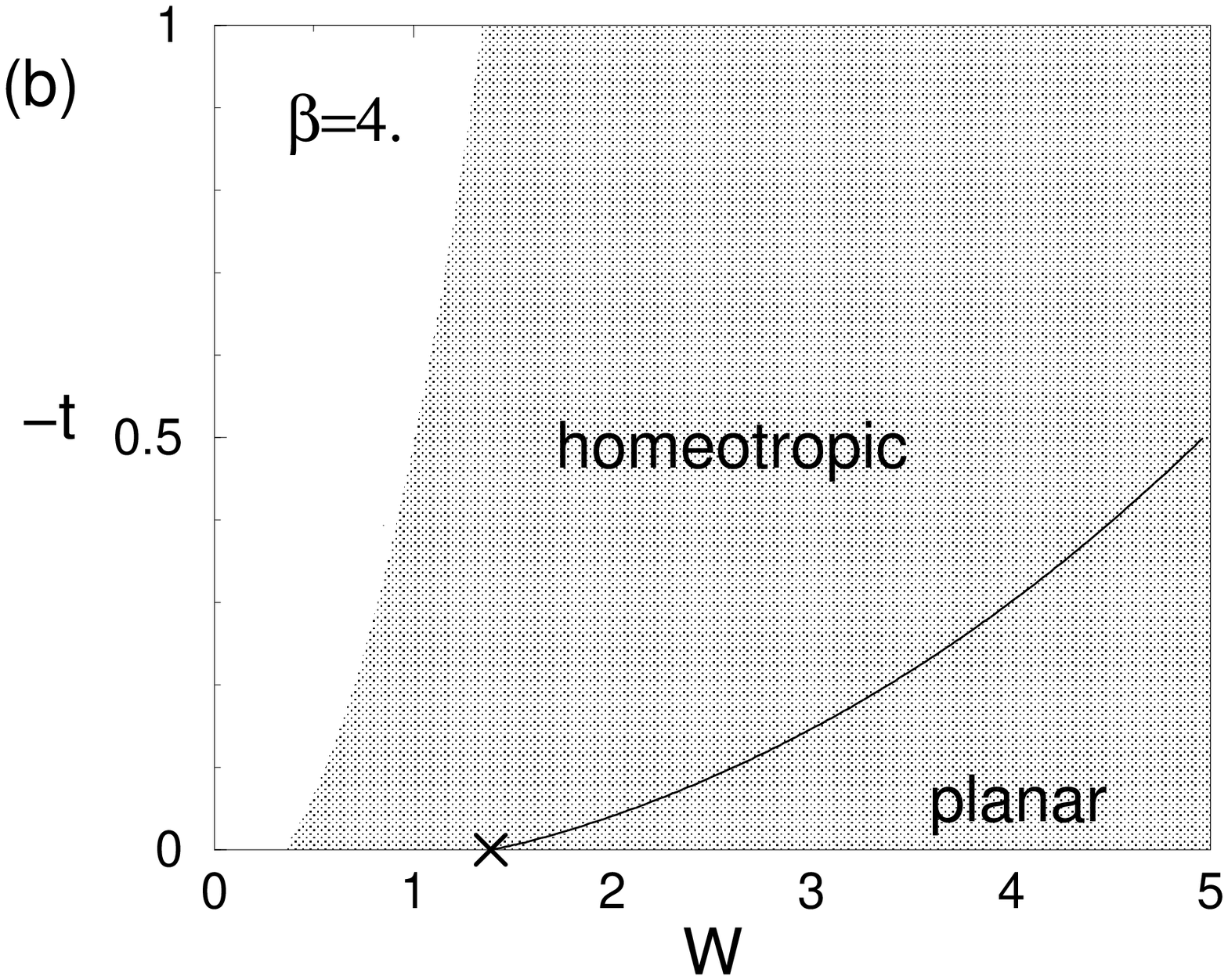}
\caption{
Anchoring phase diagrams in the $(W,t)$ plane (distance from NI coexistence $(-t)$ 
vs. surface roughness $W$), for two values of the 
substrate orienting strength $\beta$, at $\alpha = 1.2$. The solid lines mark the 
phase transition, and the hatched areas indicate the parameter regions where both 
planar and homeotropic anchoring are at least metastable. 
The cross indicates the position of the wetting transition at $t=0$.
If one increases $W$ at $t=0$ for small $\beta$ ($\beta<\beta^* = 1.16$), 
one first encounters a first order homeotropic/planar transition, and then
a second order wetting transition. For large $\beta$ ($\beta > \beta^*$), 
the homeotropic/planar transition triggers the wetting transition,
which is then first order.
}
\label{fig:transition}
\end{figure}

Some examples of anchoring potentials $\Sigma(\theta)$ are shown 
in Fig.~\ref{fig:potentials}. As a rule, the anchoring potential always 
assumed it's minimum either at $\theta = 0$ (homeotropic anchoring) or 
$\theta = \pi/2$ (planar anchoring)~\cite{fn2}.
Far from the NI coexistence and for weakly disordering substrates (low $W$), 
the orienting force of the substrate dominates and the effective anchoring is 
homeotropic. Close to the coexistence and for strongly disordering substrates 
(high $W$), the main orienting force stems from the fluid layer with strongly 
varying order parameter close to the surface. In that case, the effective anchoring 
is planar. Fig.~\ref{fig:transition} shows two examples of phase diagrams 
in the $(W,t)$ plane. At the transition lines, the surface is truly bistable, 
both homeotropic and planar alignment are equally favorable. For practical
purposes, it will often be sufficient if a state is metastable, \ie if it
corresponds to a minimum of $F(\theta)$. Fig.~\ref{fig:transition} shows 
that the regions where both states are metastable (hatched areas) are quite 
large. Moreover, the energy barriers between the two states are small 
($\sim 0.1 \hat{\epsilon}/\hat{\xi}^2 $, see Fig.~\ref{fig:potentials}), 
hence switching between the two states is easy. 

For a more thorough understanding of this transition, we must relate it to
the wetting transition. Thus we consider the two competing states,
$\theta = \pi/2$ and $\theta = 0$ at NI coexistence ($t=0$). 
The resulting free energy as a function of the surface order parameter, $S_0$, is
\begin{equation}
\Sigma = \left\{ \begin{array}{ll}
(1-S_0)^2 (1+2 S_0) + (1+\beta) \: W S_0^2  \quad      & \mbox{planar}\\
\sqrt{1+\alpha} \: (1-S_0)^2 (1+2 S_0) + W S_0^2 \quad & \mbox{homeotropic}.
\end{array}
\right.
\end{equation}
If $\Sigma$ takes it's minimum at $S_0 = 0$, the surface is wetted by the isotropic 
phase; otherwise, it is nonwet. For fixed anchoring angle, the wetting transition 
hence takes place at $W_p^* = 3/(1+\beta)$ in the planar case, and at 
$W_h^* = 3 \sqrt{1+ \alpha}$ in the homeotropic case, and it is 
continuous (critical wetting). In addition, the system may switch from the 
homeotropic state to the planar state. As expected, the wet surface 
always favors planar anchoring.

\begin{figure}
\twoimages[scale=0.28]{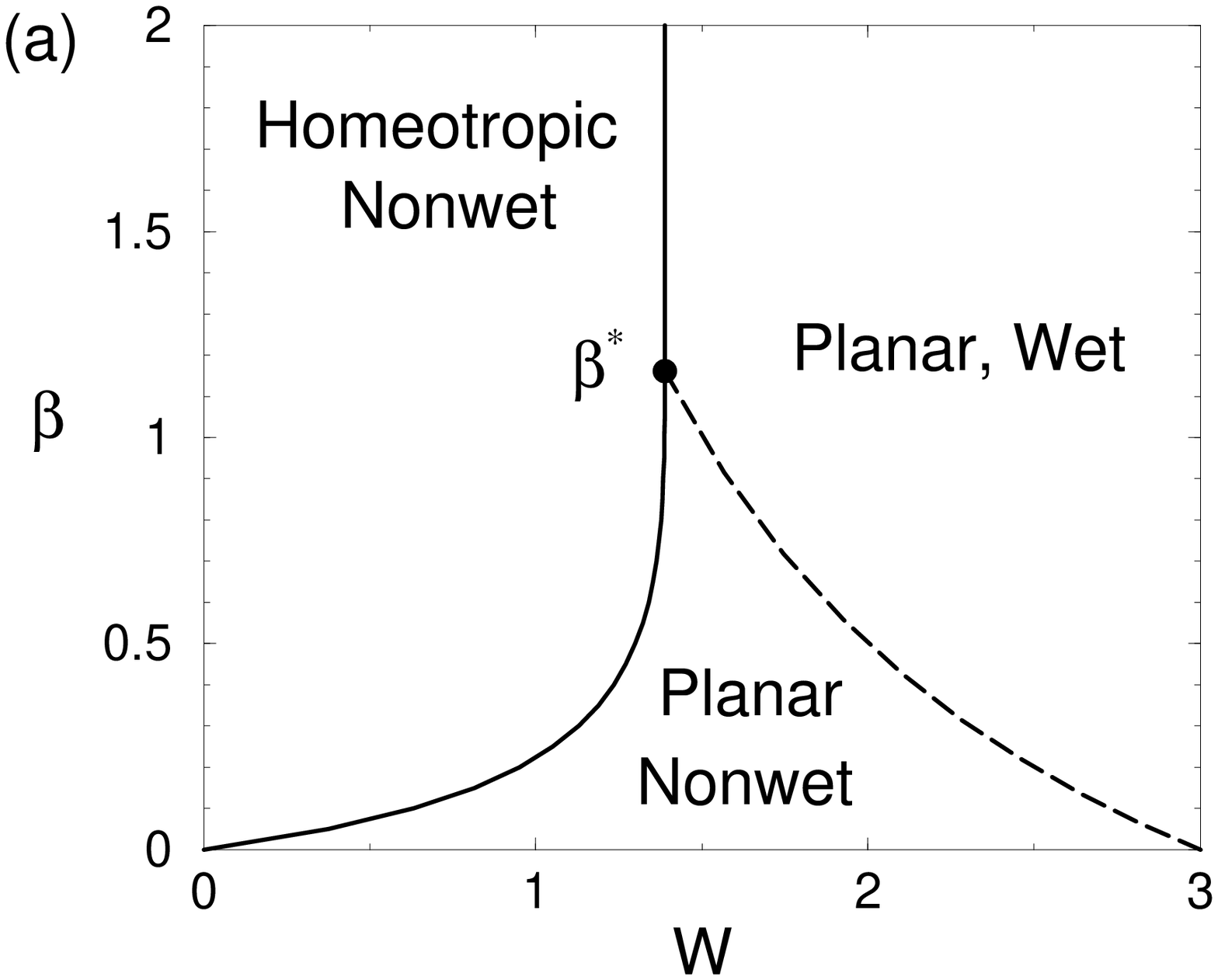}{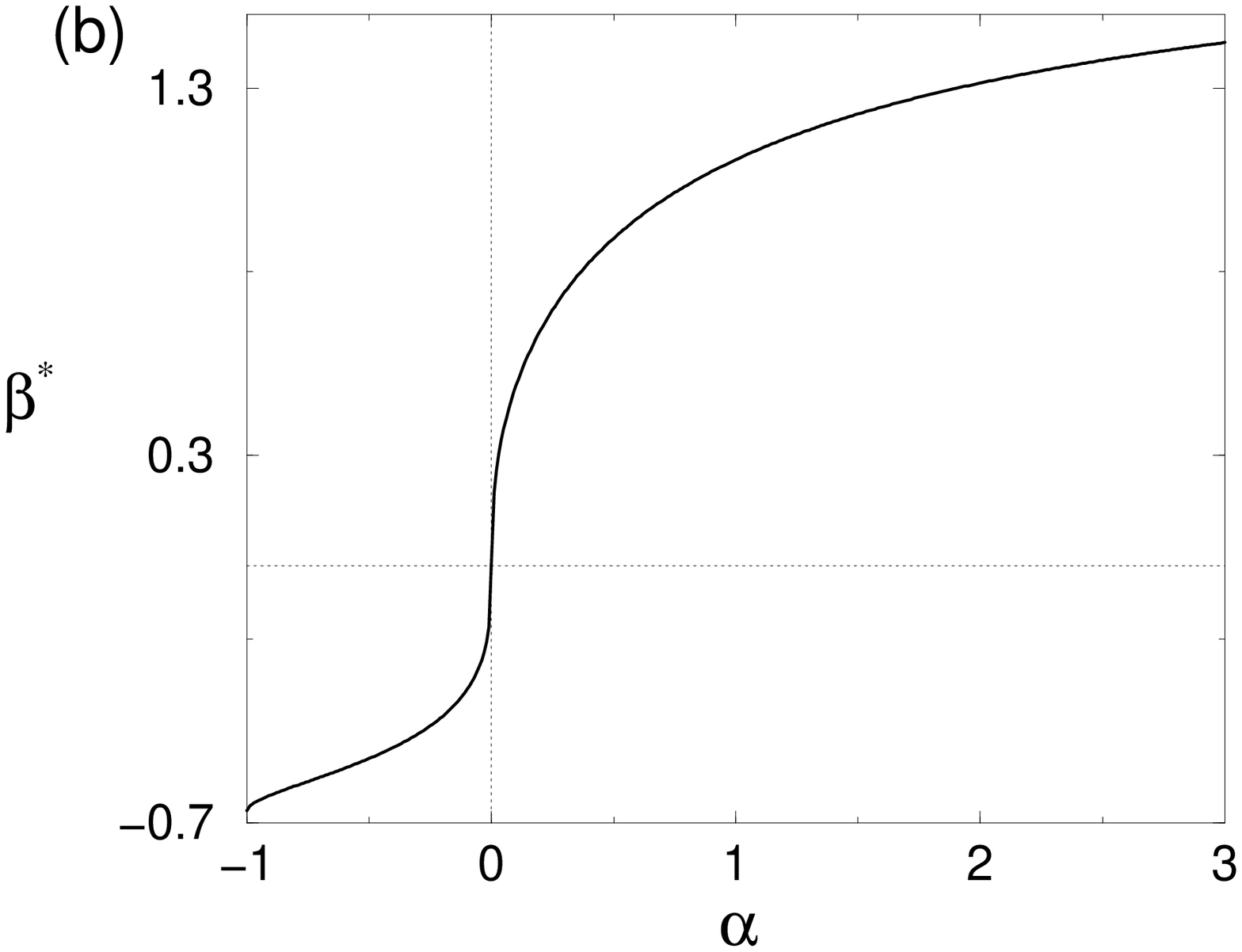}
\caption{
(a) Wetting phase diagram at coexistence ($t=0$) for $\alpha = 1.2$.
Solid lines correspond to first order, dashed line to continuous transitions.
For $|\beta| < |\beta^*|$, the planar/homeotropic phase transition 
and the wetting transition are separated; for $|\beta| > |\beta^*|$, 
they coincide. (b) Critical value $\beta^*$ where the wetting
and the anchoring transitions meet as a function of $\alpha$.
Here, results are also shown for the case $\alpha, \beta < 0$,
where the surface favors planar alignment and the NI interface 
aligns homeotropically. 
}
\label{fig:wetting}
\end{figure}

The resulting phase behavior depends on the orienting strength $\beta$ of 
the surface. If $|\beta|$ is larger than a critical value $|\beta^*|$, 
the anchoring is homeotropic for all nonwet surfaces. The transition 
from homeotropic to planar anchoring coincides with the wetting transition, 
and is first order. For smaller $|\beta|$, the (first order) 
homeotropic-planar transition preempts the wetting transition, which is 
then continuous. In this case, the system switches to planar anchoring 
at a stage where the NI interface is not yet fully developed. 
Fig.~\ref{fig:wetting} shows the wetting phase diagram for
$\alpha = 1.2$ and, more generally, the value $\beta^*$ which 
separates the two regimes as a function of $\alpha$. 

Rough, bistable, surfaces can be combined with conventional
smooth or rubbed surfaces in a nematic liquid crystal cell device.
As an example, we briefly discuss the phase diagrams of a nematic fluid 
confined between a bistable surface, and strongly orienting surfaces
with fixed anchoring angle $\theta_S = 0$ (homeotropic) or 
$\theta_S = \pi/2$ (planar). Since $S \equiv S_{\infty}$ in the cell, 
the only relevant contribution to the bulk free energy, (\ref{eq:landau}), 
is the Frank elastic energy $g_2$, and we can write the total free energy 
per area of the system as
$F_{\rm cell}/A = \Sigma(\theta_R) 
+ 3 (3 + 2 \alpha) \: S_{\infty}^2 (\theta_S - \theta_R)^2 / D.
$
Here $\theta_R$ is the anchoring angle on the rough surface, and
$D$ is the thickness of the film. This must be minimized
with respect to $\theta_R$. 
\begin{figure}
\twoimages[scale=0.28]{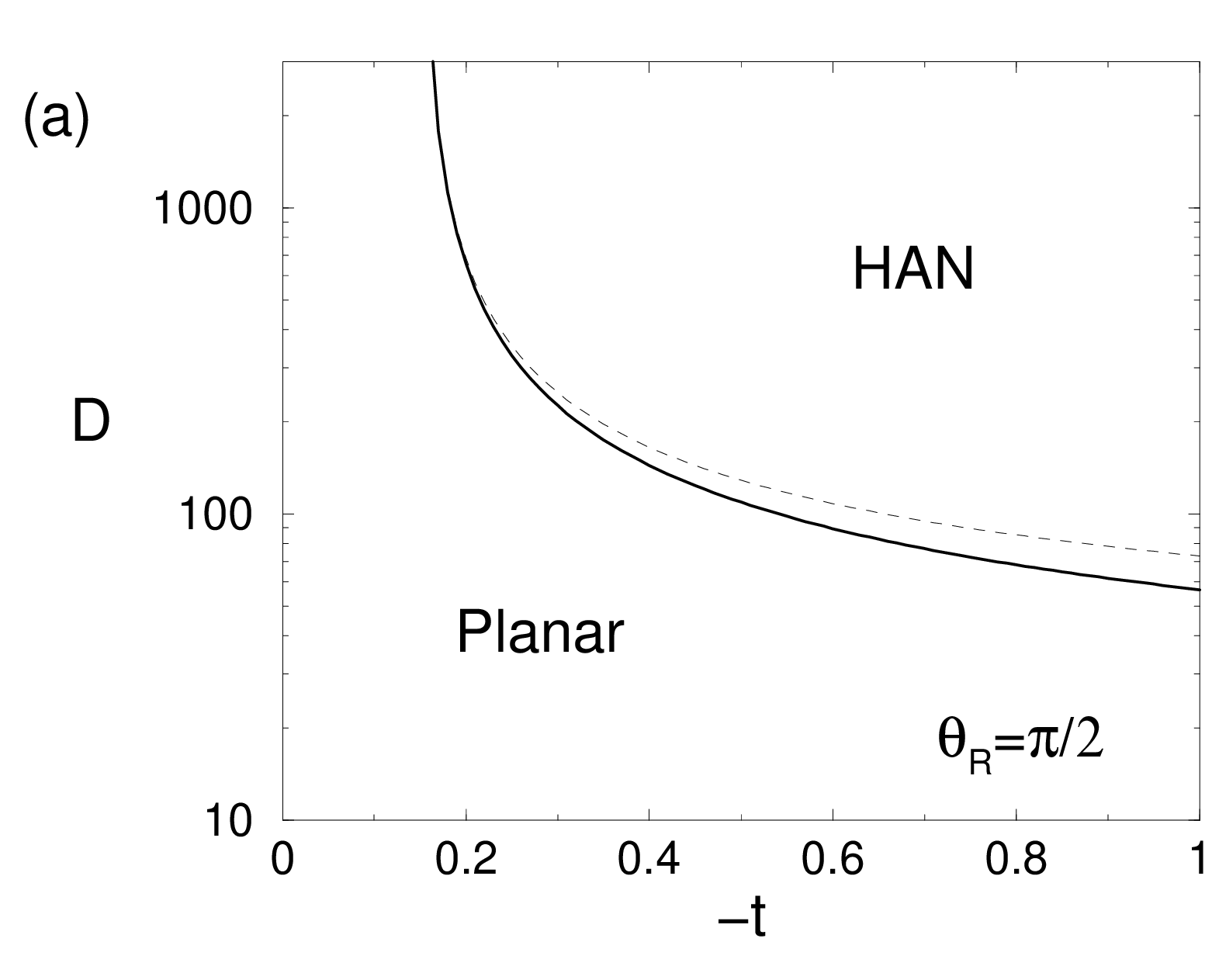}{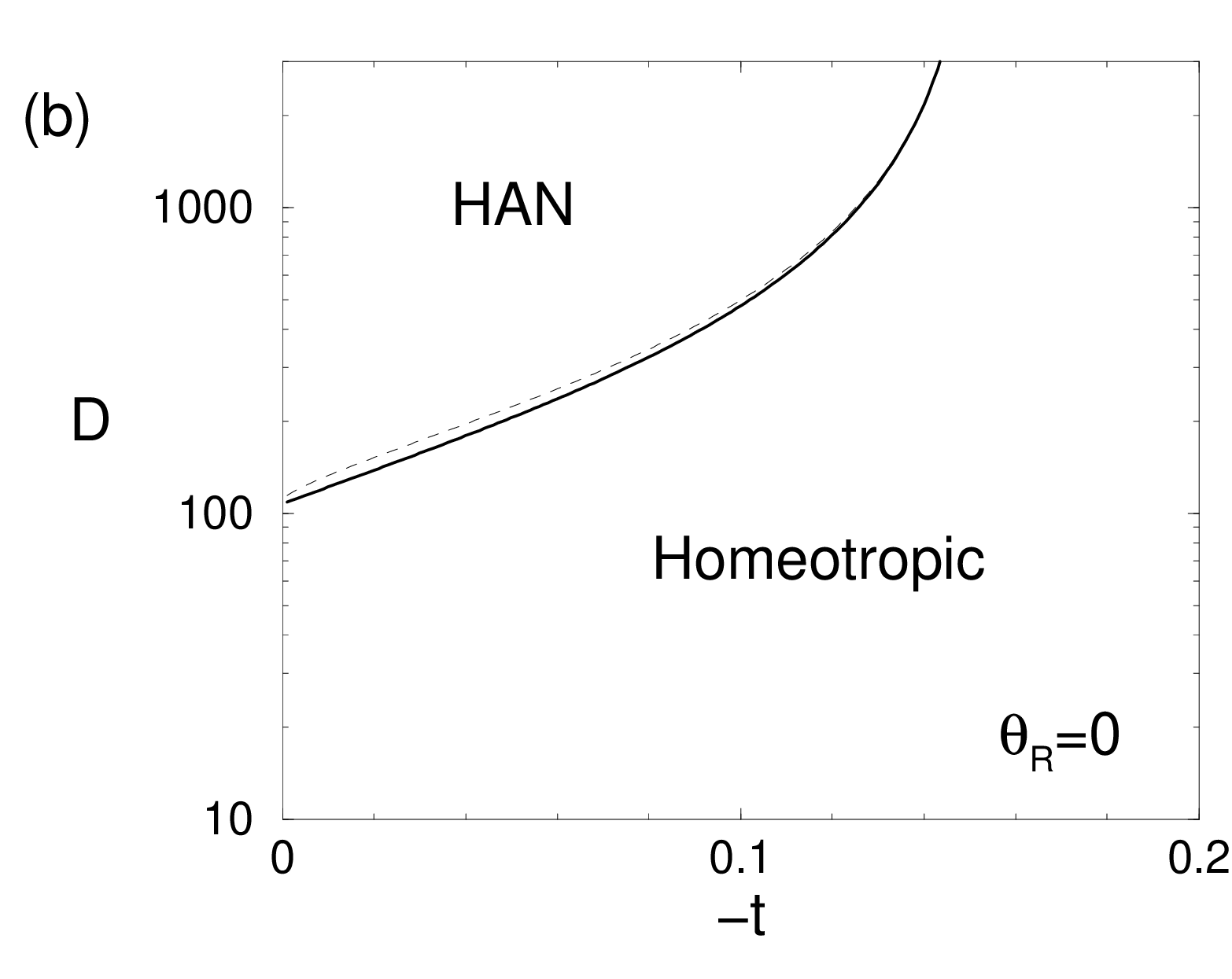}
\caption{
Phase diagrams of a nematic liquid crystal confined between a
bistable surface and a strongly anchoring surface with 
anchoring angle $\theta_R = \pi/2$ (a) and $\theta_R = 0$ (b). 
HAN denotes a ``hybrid aligned nematic'' state where the director 
rotates from homeotropic to planar alignment. The two other states 
correspond to homogeneous planar or homogeneous homeotropic
alignment. The thick solid lines indicate the exact transition 
lines, dashed lines an approximative result as explained in
the text. The parameters of the bistable surface are 
$W=2.5, \alpha = 1.2, \beta = 1.$
}
\label{fig:cell}
\end{figure}
The phase diagrams for a system with the same surface parameters as in 
Fig.~\ref{fig:potentials} are shown in Fig.~\ref{fig:cell}. The system 
can assume two different configurations - a homogeneous configuration 
with fixed director angle, $\theta \equiv \theta_R$ throughout the system, 
and a hybrid aligned nematic (HAN) configuration where the director slowly 
rotates from the planar to the homeotropic orientation in the $z$-direction.
The transition lines between the different states can be calculated 
to a very good approximation, if one assumes that $\theta_R$ 
can only take the values $\theta_R=0$ or $\theta_R = \pi/2$ (dashed lines
in Fig.~\ref{fig:cell}). Hence the rough surface really acts like
a two-state surface.

To summarize, we have analyzed the interplay of wetting and anchoring 
on disordering substrates within the Landau-de Gennes theory in situations
of competing, homeotropic and planar, anchoring axes at the substrate
and the NI interface. We have identified different substrate regimes -- 
one ``strongly orienting'' regime where the anchoring transition coincides 
with the wetting transition, and one ``weakly orienting'' regime where it 
preempts the wetting transition. The anchoring transition is 
first order, but the wetting transition must be continuous, if it is not 
coupled with an anchoring transition.  Finally, we have discussed how this 
effect can be used to design bistable surfaces, which favor two distinctly 
different orientations, and given an example how such a surface could be 
integrated in a nematic liquid crystal cell. 
The present study is a mean field study, the effect of fluctuations has been 
disregarded. Two types of fluctuations will renormalize the surface potential: 
Director fluctuations become increasingly important close to the NI 
transition~\cite{fournier}, and fluctuations of the NI 
interface~\cite{elgeti} become important at complete wetting. It will be
interesting to study the influence of these effects in future work.

This work was funded by the DFG and the EPSRC.


\begin{thebibliography}{0}
\bibitem{fn1} For a first order wetting transition, one would
need a surface term which is proportional to $S$.
\bibitem{fn2} This also seems to hold for nonzero $\gamma$, except
in cases where the homeotropic anchoring is dominated by the fourth
order term, $\gamma \gg \beta \approx 0.$ The anchoring on the high 
$|t|$ side can then become conical. In the inverse case 
$\alpha,\beta < 0, \gamma =0$ (planar anchoring at the substrate, 
homeotropic anchoring at the NI interface), a weak metastable 
minimum at $\theta = \pi/4$ is found in a very narrow parameter 
range at $\gamma = 0$. Note that for $\beta < 0$, $\gamma > 0$, 
the surface eventually favors conical anchoring (as in the
system of Ref.~\protect\cite{teixeira}).
\end{thebibliography}
\end{document}